\documentclass[a4paper,preprint,superscriptaddress,preprintnumbers,nofootinbib]{revtex4}
\usepackage{amsmath,amssymb}
\usepackage{bm}
\usepackage[dvipdfmx]{graphicx}
\newcommand{\Slash}[1]{{\ooalign{\hfil/\hfil\crcr$#1$}}}
\DeclareMathOperator*{\p}{\mathcal{D} \psi \mathcal{D} \bar{\psi}}

\begin{document}

\title{Dynamical chiral symmetry breaking and \\
weak nonperturbative renormalization group equation \\
in gauge theory}

\author{Ken-Ichi \surname{Aoki}}
\email{aoki@hep.s.kanazawa-u.ac.jp}
\affiliation{Institute for Theoretical Physics, Kanazawa University, Kanazawa 920-1192, Japan}

\author{Shin-Ichiro \surname{Kumamoto}}
\email{kumamoto@hep.s.kanazawa-u.ac.jp}
\affiliation{Institute for Theoretical Physics, Kanazawa University, Kanazawa 920-1192, Japan}

\author{Daisuke \surname{Sato}\footnote{until September 2014}}
\email{satodai@hep.s.kanazawa-u.ac.jp}
\affiliation{Institute for Theoretical Physics, Kanazawa University, Kanazawa 920-1192, Japan}

\preprint{KANAZAWA-16-06}

\begin{abstract}
We analyze the dynamical chiral symmetry breaking (D$\chi$SB) in gauge theory with the nonperturbative renormalization group equation (NPRGE), which is a first order nonlinear partial differential equation (PDE). In case that the spontaneous chiral symmetry breaking occurs, the NPRGE encounters some non-analytic singularities at the finite critical scale even though the initial function is continuous and smooth. Therefore there is no usual solution of the PDE beyond the critical scale. 
In this paper, we newly introduce the notion of a weak solution which is the global solution of the weak NPRGE. We show how to evaluate the physical quantities with the weak solution.
\end{abstract}

\maketitle

\section{Introduction}
There are various methods to analyze the D$\chi$SB.
In particular, the nonperturbative renormalization group is quite effective since it may include non-ladder diagrams which cure the gauge invariance problem \cite{Miya09, Sato13}.
However, it is difficult to solve the NPRGE by the usual difference method and the power expansion in operators,
since the solution has some singularities such as discontinuity and non-differentiability when D$\chi$SB occurs.
Accordingly, we introduce a new method to analyze the D$\chi$SB by the weak NPRGE in a gauge theory with a non-running gauge coupling constant.

\section{Nonperturbative renormalization group equation}
The partition function is given by the path integral in the momentum representation, 
\begin{equation}
Z =  \int \p_{0 \le |p| \le \Lambda}  \exp(-S_{\Lambda}),  
\end{equation}
where $p$, $\Lambda$ and $S_{\Lambda}$ are the momentum of  fields, the cutoff and the Wilsonian effective action, respectively. 
As the initial condition 
at the ultraviolet cutoff $\Lambda_0$, we take $S_{\Lambda_0} =\int d^4 x ~\mathcal{L_{\rm  gauge}}$. 
If the cutoff reaches down to 0, that is, the infrared limit, we obtain $S_{0} =-\log Z$ which contains all quantum effects.
We adopt the local potential approximation for gauge theory 
and the Wilsonian effective action reads
\begin{equation}
S_{\Lambda (t) } \simeq \int d^4 y 
~[{\bar{\psi}} ( i \Slash{\partial} +g \Slash{A}){\psi}
+\frac{1}{4}{F_{\mu \nu}}^2+\frac{1}{2\xi}(\partial_{\mu} A_{\mu})^2 -   V_{\rm W}({\bar{\psi}} {\psi},t)  ],
\label{eq:LPA} 
\end{equation}
where $g$, $\xi$ are the gauge coupling constant and the gauge fixing parameter, respectively.
We denote $\Lambda(t) \equiv \exp(-t) \Lambda_0$, and $t$ is the renormalization group scale parameter. 
The last term $V_{\rm W}({\bar{\psi}} {\psi},t)$ in Eq.\,(\ref{eq:LPA}) is the Wilsonian effective potential, which consists of fermion fields only.
Then we obtain the first order nonlinear PDE, the so-called  nonperturbative renormalization group equation \cite{Weg73},
\begin{equation}
\frac{\partial V_{\rm W}(x,t)}{\partial t} +f(x,M,t) =0, 
~ x \equiv {\bar{\psi}} {\psi},
~M(x,t)  \equiv   \frac{\partial V_{\rm W} (x,t)}{\partial x}.
\label{eq:HJE} 
\end{equation}
The so-called flux function $f(x,M,t)$ for gauge theory is given by
\begin{equation}
f(x,M,t) \equiv -   \frac{e^{-4t}}{4\pi^2} \log \left[ 1+(M+b(t) x)^2 \right]
,~  b(t) \equiv \frac{(3+\xi) \pi {g}^2 {\rm e}^{2t}}{4 \pi^2},~ \xi=0.
\end{equation}
This PDE has the same form as the Hamilton-Jacobi equation appearing in analytical mechanics.
We can interpret $x$, $t$, $M$, $V_{\rm W}(x,t)$ and $f(x,M,t)$ as position, time, momentum, action and Hamiltonian, respectively.
We solve this PDE with the initial condition $V_{\rm W}(x,0)=m_0 x$ coming from the bare mass term $m_0 \bar{\psi} {\psi} $ in the Lagrangian.

\section{The method of characteristics}
Any first order PDE can be deformed to a system of ordinary differential equations, the so-called characteristic equations.
For the Hamilton-Jacobi equation the characteristic equations are nothing but the Hamilton's canonical equations. 
The characteristic equations for NPRGE (\ref{eq:HJE}) are obtained as
\begin{equation}
\begin{cases}
& \displaystyle \frac{d x_{\rm c} (t)}{d t}=\left. \frac{\partial f(x,M,t  ) }{\partial M} \right|_{\substack{{x=x_{\rm c}(t)}\\{M=M_{\rm c}(t)}}},  \vspace{1mm} \\ 
& \displaystyle  \frac{d M_{\rm c}(t) }{d t}=- \left. \frac{\partial f(x,M,t) }{\partial x} \right|_{\substack{{x=x_{\rm c}(t)}\\{M=M_{\rm c}(t)}}}. 
\end{cases}
\end{equation}
Then we can calculate the mass function $M(x,t)$ at any scale $t$ as Fig.\,1 (a).
Though the mass function is the multi-valued function beyond the critical scale $t_{\rm c}$, it must be the single-valued function  at any scale because it is the physical quantity corresponding to the physical effective theory at the scale.
Therefore we introduce the {\it weak solution}, the mathematically extended notion of solution which is a single-valued solution and can have some discontinuous points.

\def\figsubcap#1{\par\noindent\centering\footnotesize(#1)}
\begin{figure}
\begin{center}
  \parbox{2.5in}{\includegraphics[width=2.5in]{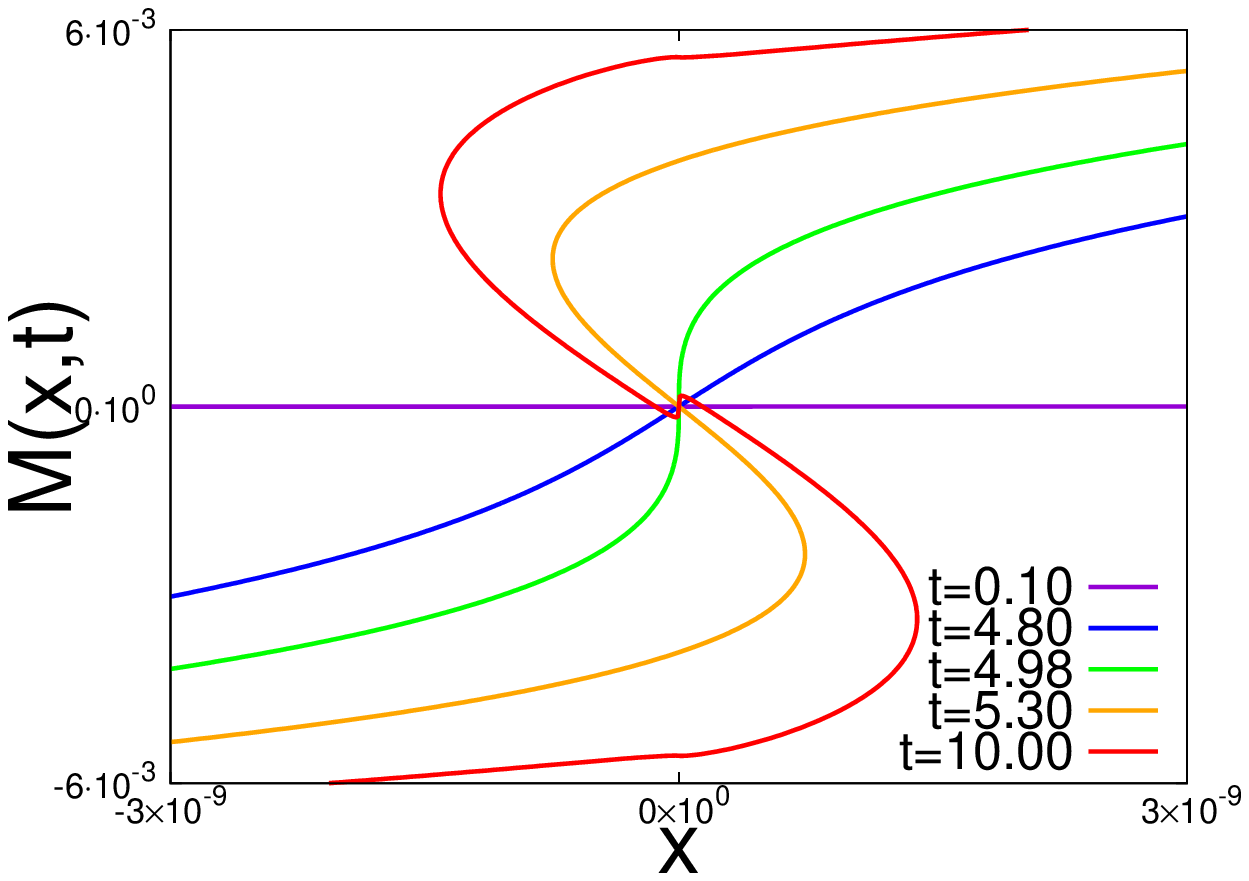}}
  \hspace*{20pt}
  \parbox{1.5in}{\includegraphics[width=1.5in]{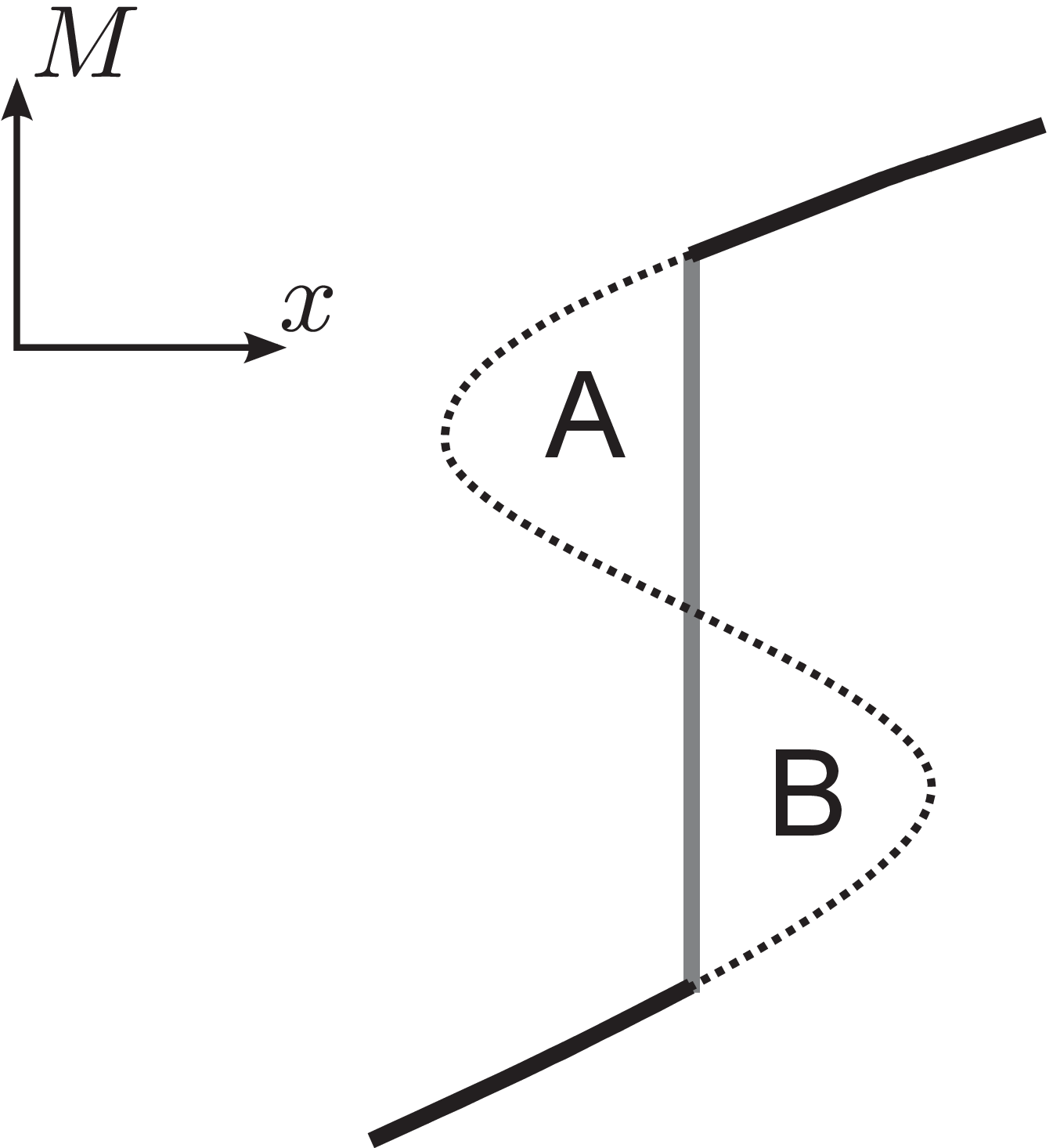}}
  \caption{(a) Mass function ($g=1.1 g_{\rm c}$). (b) Equal area rule.}%
  \label{fig1}
\end{center}
\vskip-0.5cm
\end{figure}

\section{ Weak solution}
There are two major methods to define weak solutions.
One is to use the Hamilton-Jacobi equation \cite{Cl-Li81, Cl-Li83}, and the other is to use the conservation law type equation \cite{Bur40} which is obtained by differentiating the Hamilton-Jacobi equation with respect to $x$.  
We adopt the latter in this paper.
The conservation law for our system is written as
\begin{equation}
\frac{\partial M(x,t) }{\partial t} +   \frac{\partial f(x,M,t) }{\partial x}  =0, 
~M(x,0)=m_0.
\label{eq:CL} 
\end{equation}
Then we multiply Eq.\,(\ref{eq:CL}) by an arbitrary test function $\varphi(x,t)$ which is  continuously  differentiable and vanishes at $t=\infty$ and $x=\pm \infty$, and integrate it with respect to $x$ and $t$.
Thus we obtain the integral equation, 
\begin{equation}
\int_{0}^{\infty}  dt  \int_{- \infty}^{\infty}  dx  \left[ ~\frac{\partial  M(x,t)}{\partial t}+\frac{\partial f(x,M,t) }{\partial x}   \right] \varphi(x,t)=0.
\end{equation}
Then we perform the integration by parts and obtain the weak NPRGE,
\begin{equation}
\int_{0}^{\infty}  dt   \int_{- \infty}^{\infty}  dx   \left[ M(x,t)  \frac{\partial  \varphi(x,t)}{\partial t} +f(x,M,t) \frac{\partial  \varphi(x,t)}{\partial x} \right] +  \int_{- \infty}^{\infty}  dx  ~M(x,0) \varphi (x,0) =0.
\label{eq:WNPRGE} 
\end{equation}
Function $M(x,t)$ which fulfills Eq.\,(\ref{eq:WNPRGE}) for any test function is called the weak solution of the original PDE (\ref{eq:CL}) and this $M(x,t)$ is no longer required to be continuous.
The graphical expression of the weak solution is simply the equal area rule of the mass function \cite{Whitham74}.
Namely all we have to do is to select the position of discontinuity so that the area A is equal to B in Fig.\,1 (b), which resembles to the Maxwell construction.

\section{Physical interpretation of the weak solution and summary}
In Fig.\,2 we show the scale $t$ evolution of physical quantities, the mass function $M(x,t)$, the Wilsonian effective potential $V_{\rm W}(x,t)$ and the Legendre effective potential $V_{\rm L}(\varphi,t)$. 
Broken lines are the results of Hamilton's canonical equations, and solid lines are the weak solution. 
The weak solution of the NPRGE convexifies the Legendre effective potential, which assures the weak solution is correct physically.

This method to use the weak NPRGE works well for extended theories, for example, QCD at finite temperature and density, and even beyond the local potential approximation.

\clearpage

\begin{figure}[h]
  \begin{center}
  \vspace{-10mm}
    \begin{tabular}{c}   
      \hspace{5mm}
      \begin{minipage}{0.3\hsize}
        \begin{center}
          \includegraphics[width=5cm]{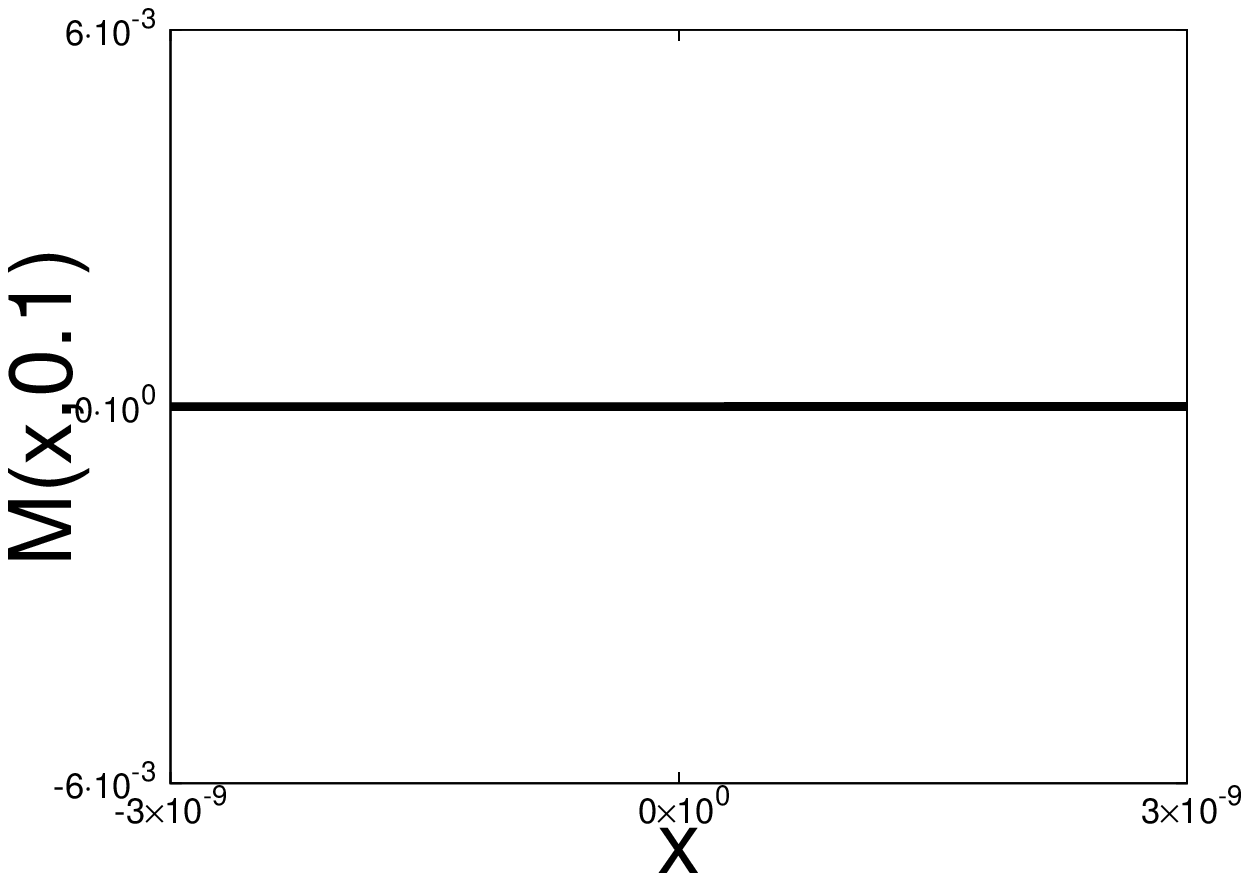}
          
          \vspace{-2mm}\hspace{5mm}(1a)\vspace{5mm}
        \end{center}
      \end{minipage}
      \hspace{2mm}
      \begin{minipage}{0.3\hsize}
        \begin{center}
          \includegraphics[width=5cm]{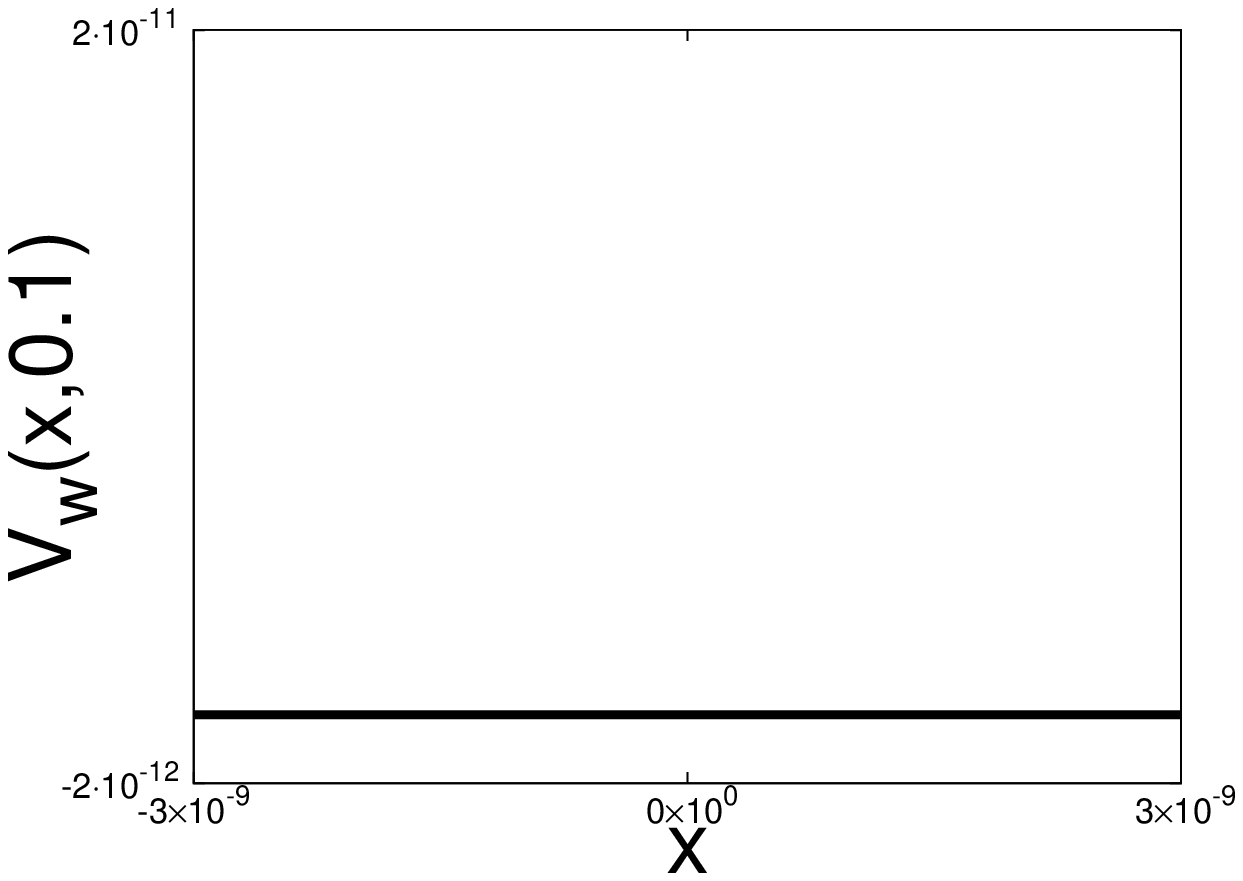}
          
          \vspace{-2mm}\hspace{5mm}(1b)\vspace{5mm}
        \end{center}
      \end{minipage}
      \hspace{2mm}
      \begin{minipage}{0.3\hsize}
        \begin{center}
          \includegraphics[width=5cm]{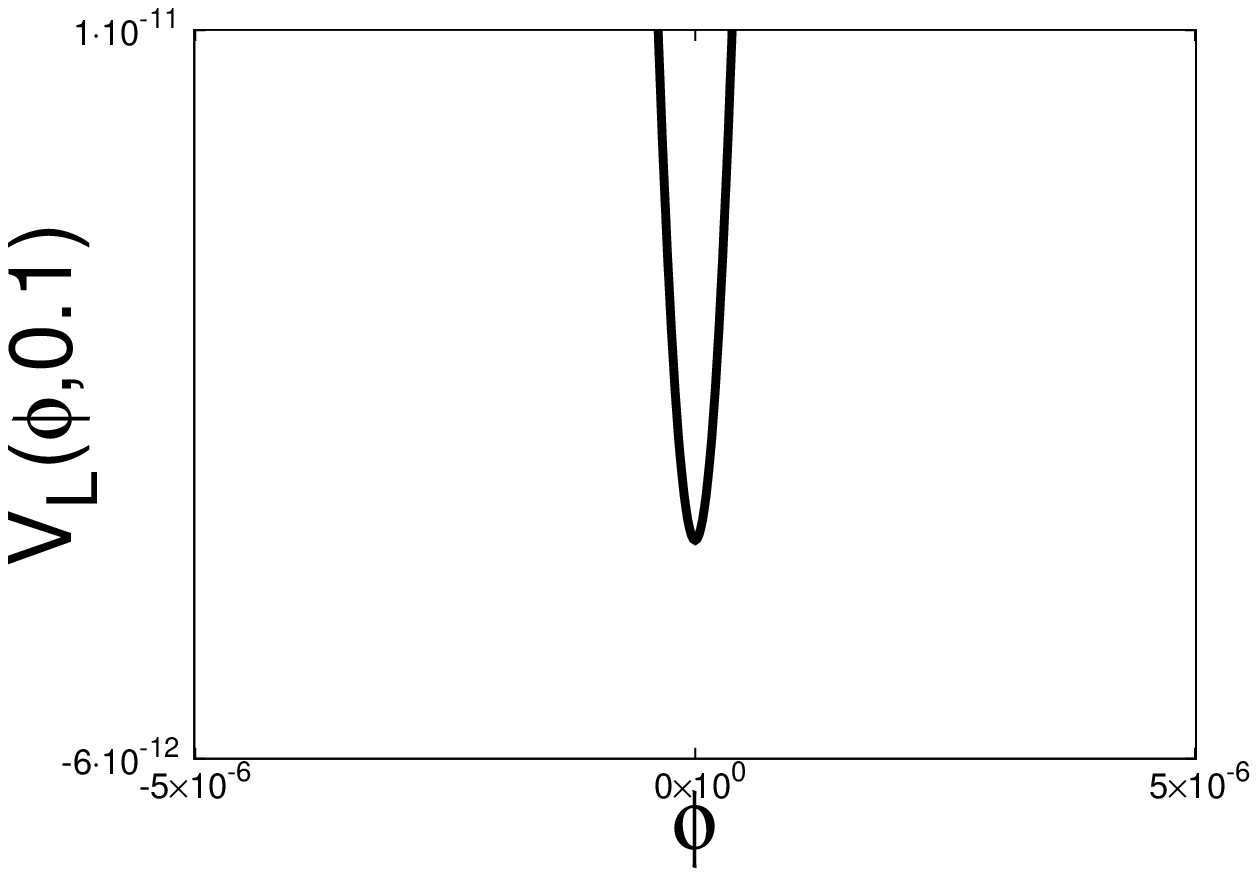}
          
          \vspace{-2mm}\hspace{5mm}(1c)\vspace{5mm} 
        \end{center}
      \end{minipage}
    \end{tabular}
    \begin{tabular}{c}
      \hspace{5mm}
      \begin{minipage}{0.3\hsize}
        \begin{center}
          \includegraphics[width=5cm]{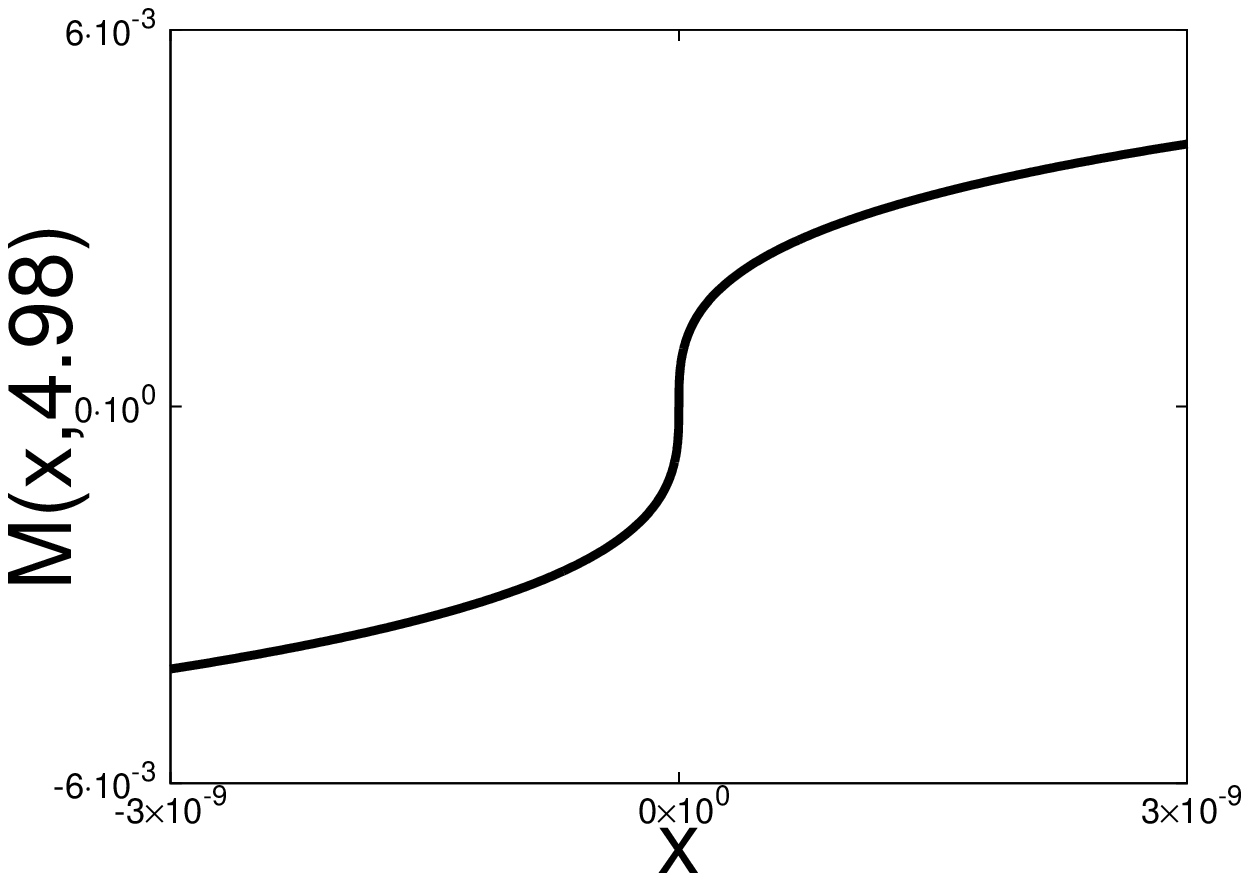}
         
         \vspace{-2mm}\hspace{5mm}(2a)\vspace{5mm}
        \end{center}
      \end{minipage}
      \hspace{2mm}
      \begin{minipage}{0.3\hsize}
        \begin{center}
          \includegraphics[width=5cm]{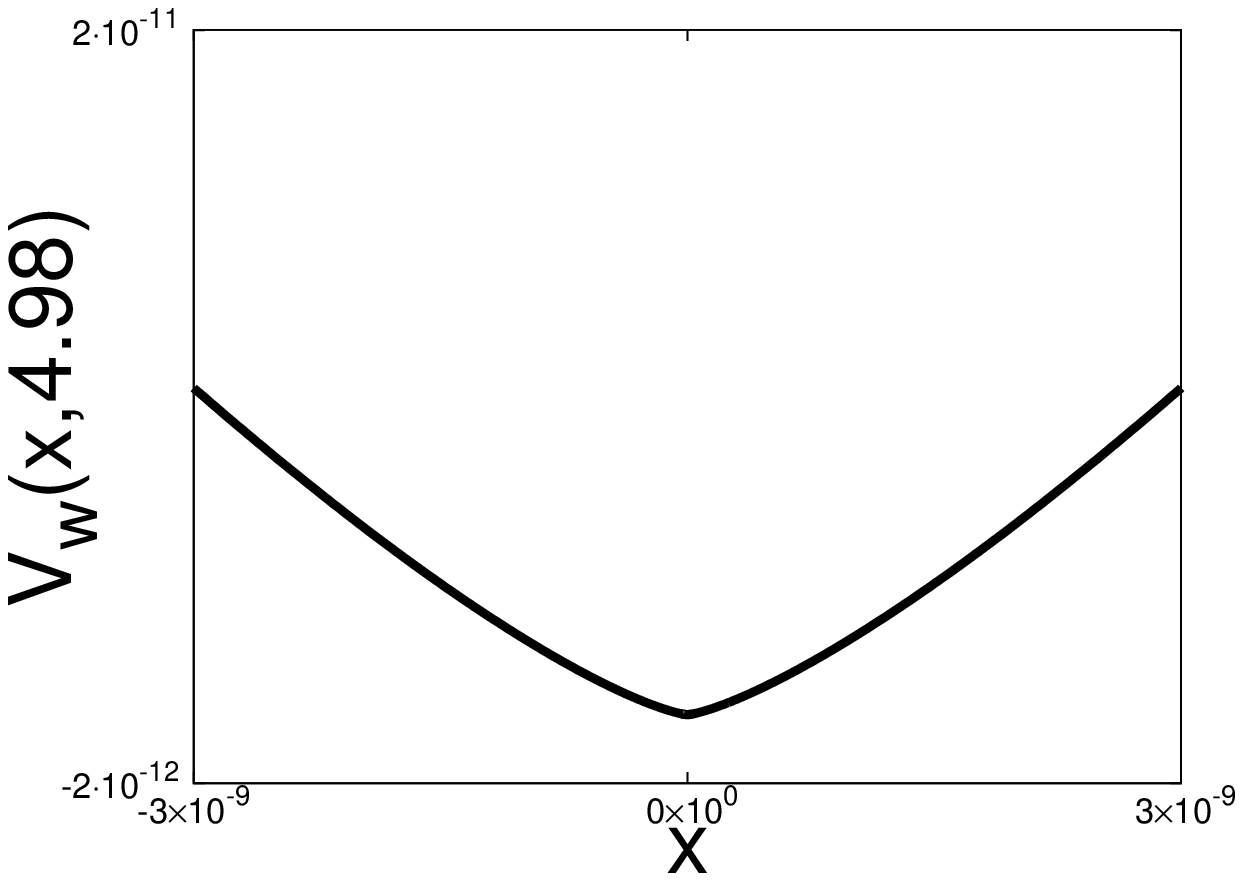}
          
          \vspace{-2mm}\hspace{5mm}(2b)\vspace{5mm}
        \end{center}
      \end{minipage}
      \hspace{2mm}
      \begin{minipage}{0.3\hsize}
        \begin{center}
          \includegraphics[width=5cm]{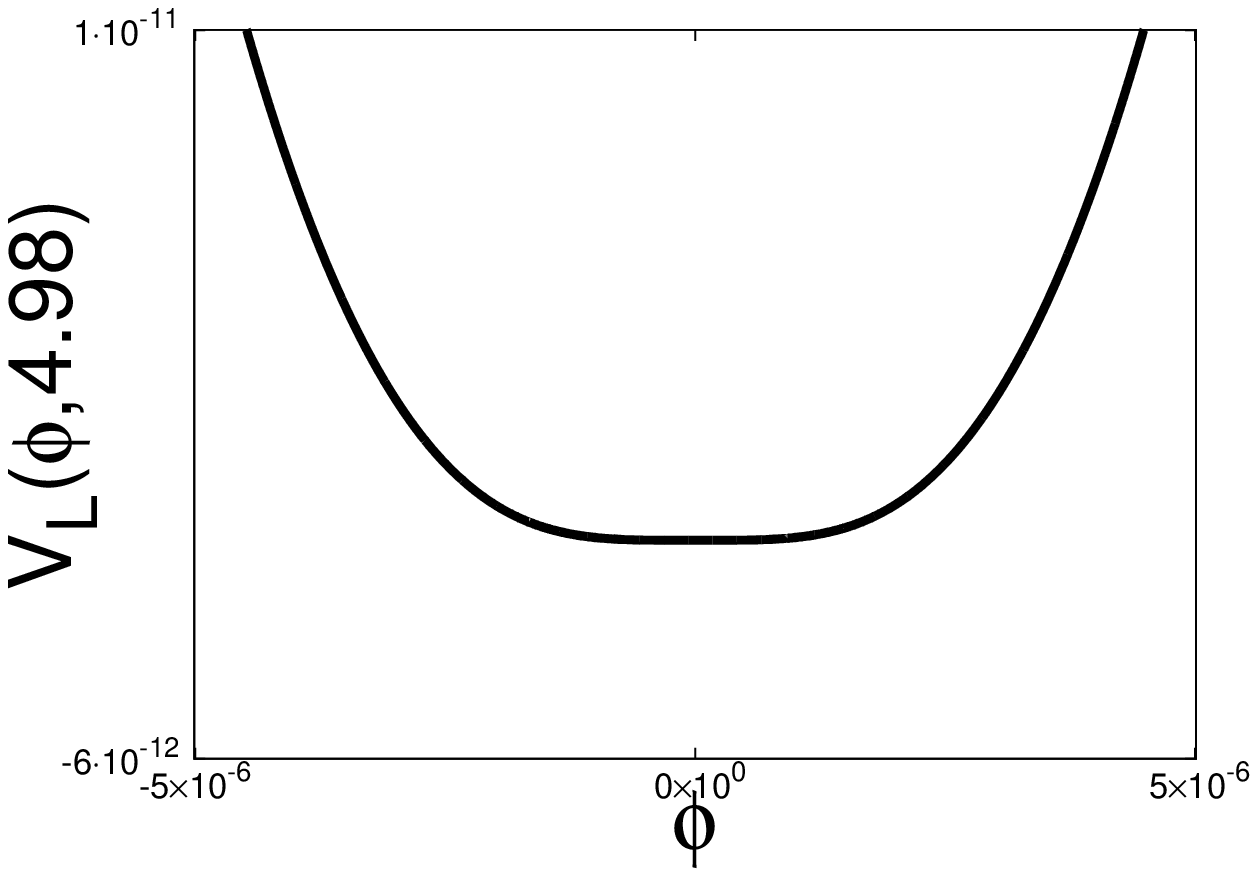}
          
          \vspace{-2mm}\hspace{5mm}(2c)\vspace{5mm}
        \end{center}
      \end{minipage}
    \end{tabular}      
    \begin{tabular}{c}
      \hspace{5mm}
      \begin{minipage}{0.3\hsize}
        \begin{center}
          \includegraphics[width=5cm]{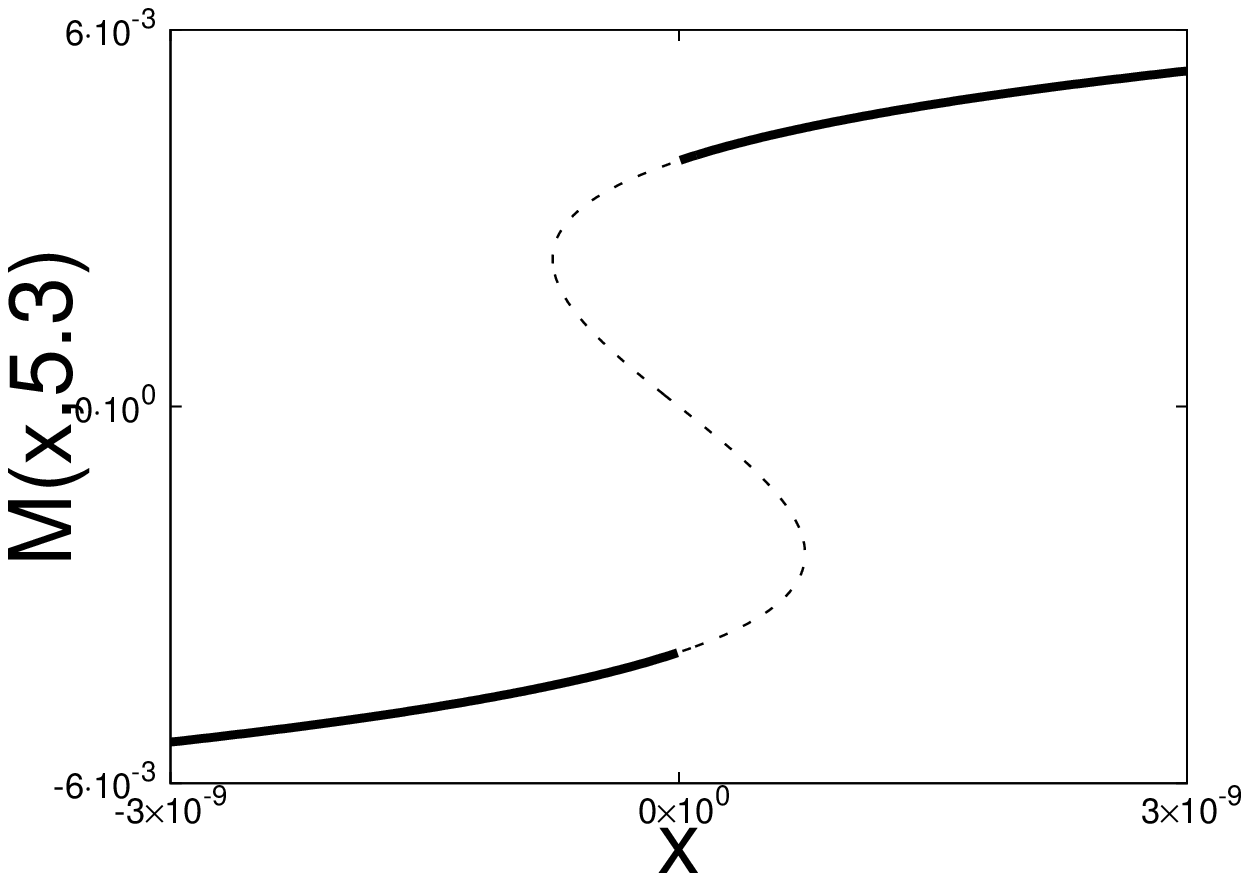}
         
         \vspace{-2mm}\hspace{5mm}(3a)\vspace{5mm}
        \end{center}
      \end{minipage}
      \hspace{2mm}
      \begin{minipage}{0.3\hsize}
        \begin{center}
          \includegraphics[width=5cm]{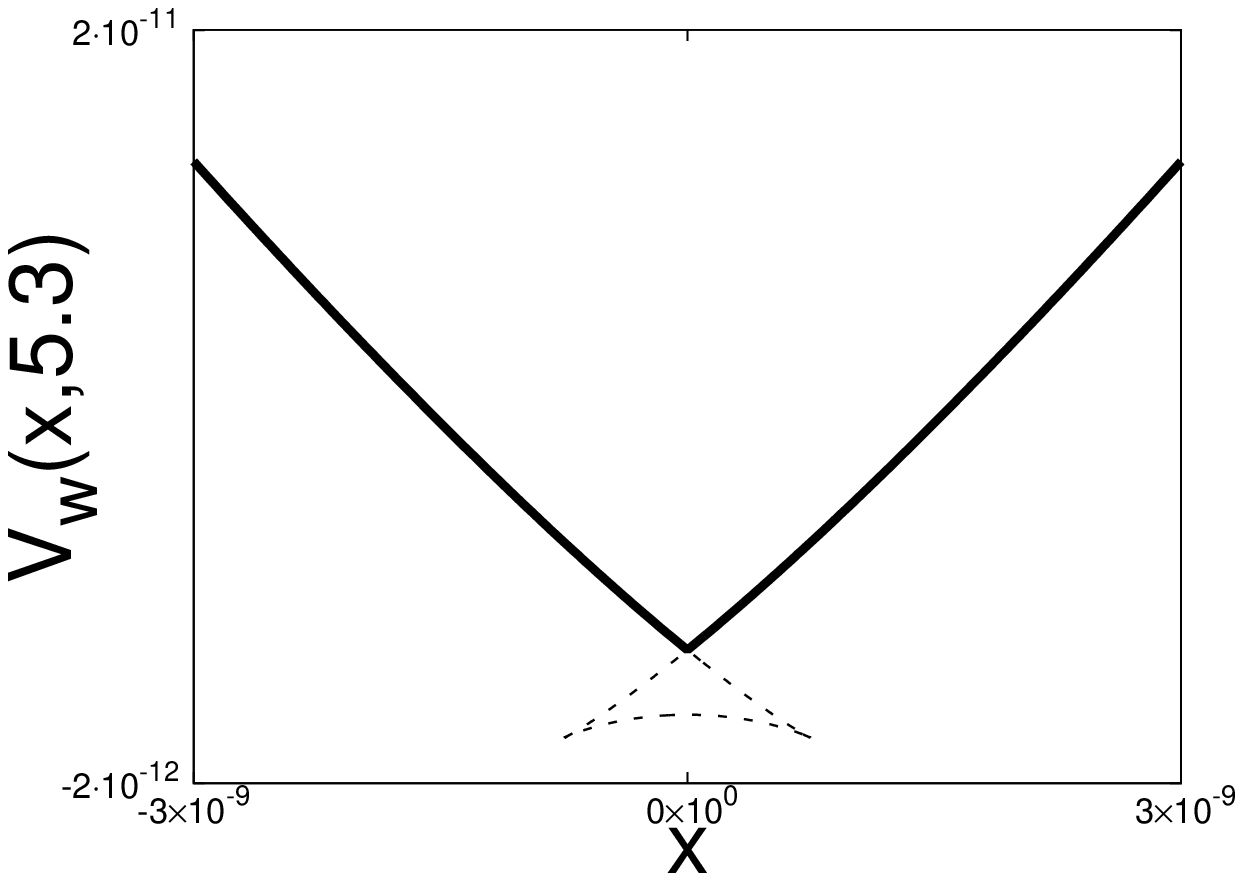}
         
         \vspace{-2mm}\hspace{5mm}(3b)\vspace{5mm}
        \end{center}
      \end{minipage}
      \hspace{2mm}
      \begin{minipage}{0.3\hsize}
        \begin{center}
          \includegraphics[width=5cm]{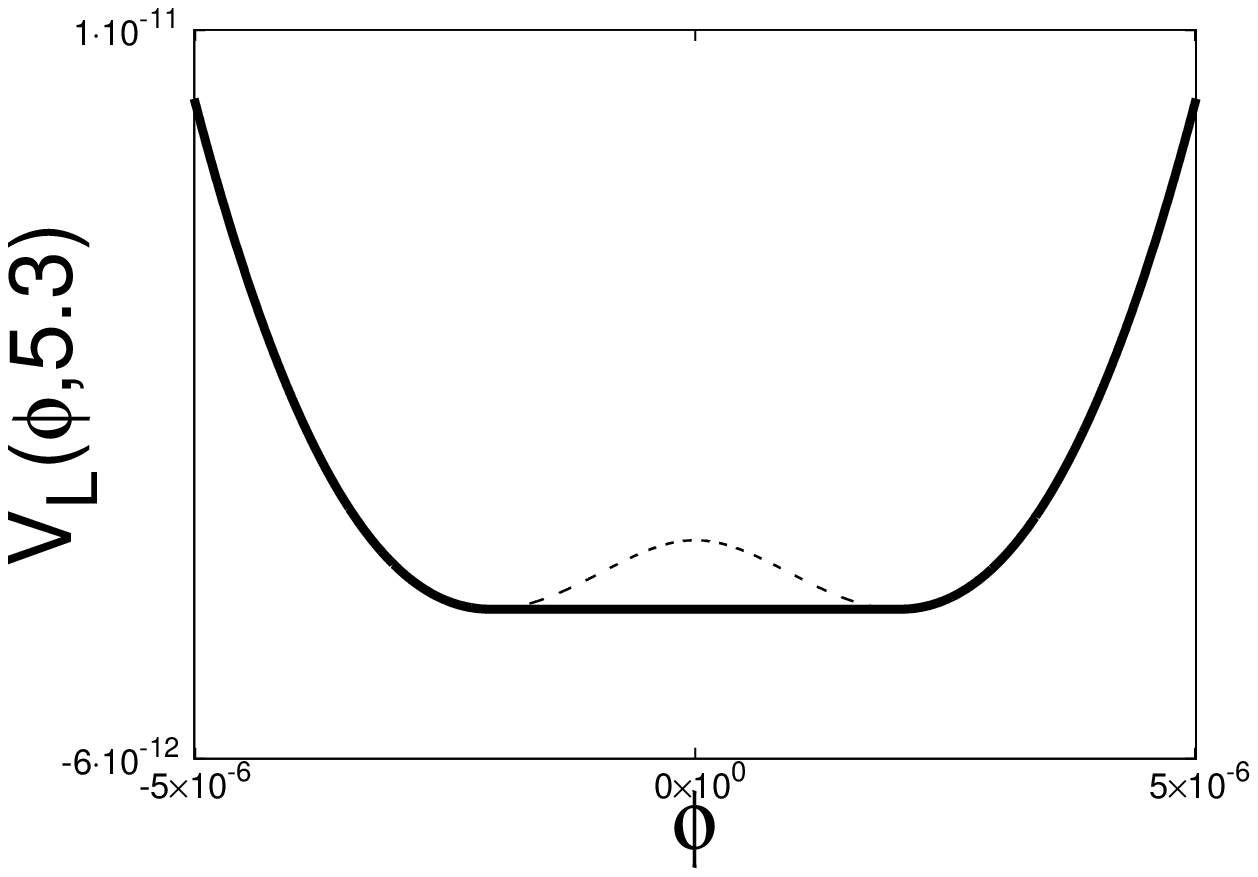}
          
          \vspace{-2mm}\hspace{5mm}(3c)\vspace{5mm}
        \end{center}
      \end{minipage}
    \end{tabular}
    \begin{tabular}{c}
      \hspace{5mm}
      \begin{minipage}{0.3\hsize}
        \begin{center}
          \includegraphics[width=5cm]{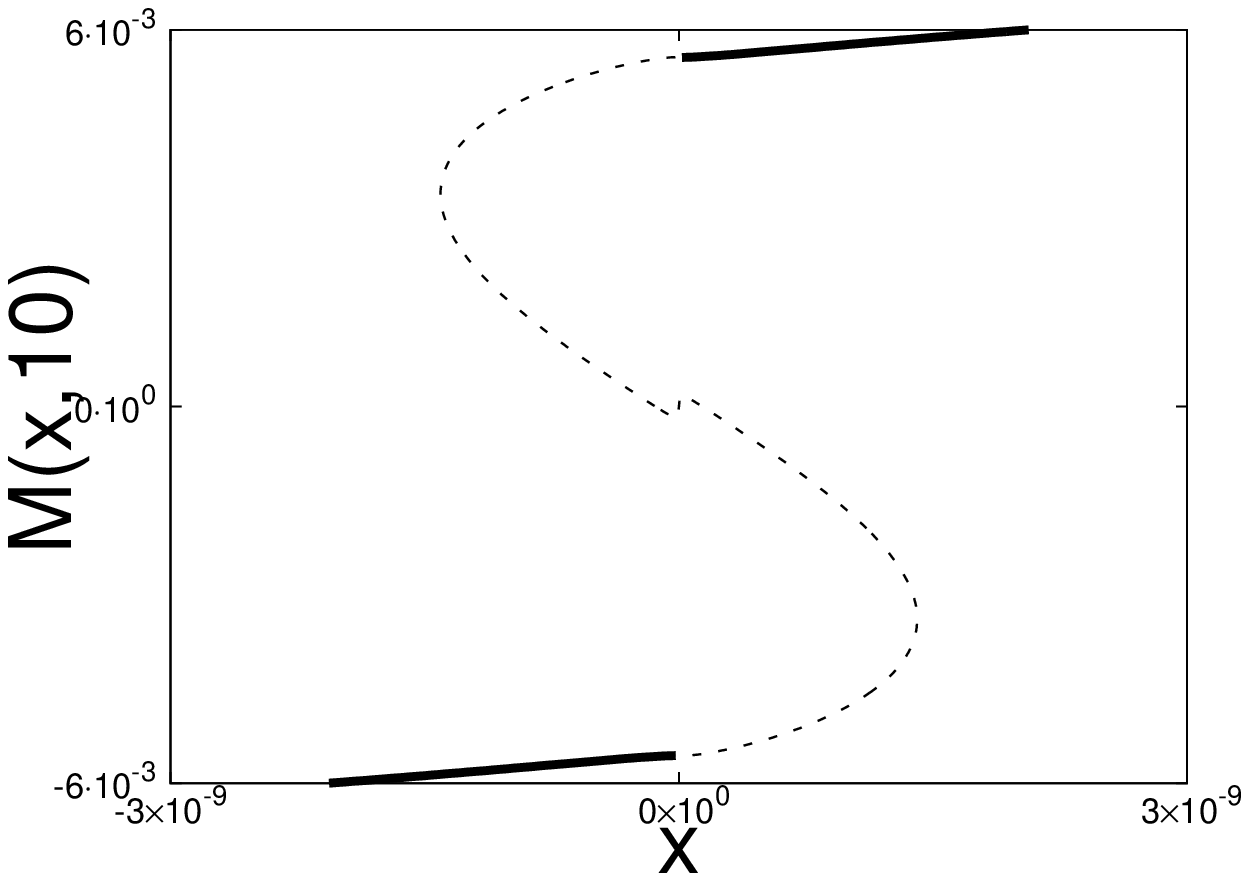}
         
         \vspace{-2mm}\hspace{5mm}(4a)\vspace{5mm}
        \end{center}
      \end{minipage}
      \hspace{2mm}
      \begin{minipage}{0.3\hsize}
        \begin{center}
          \includegraphics[width=5cm]{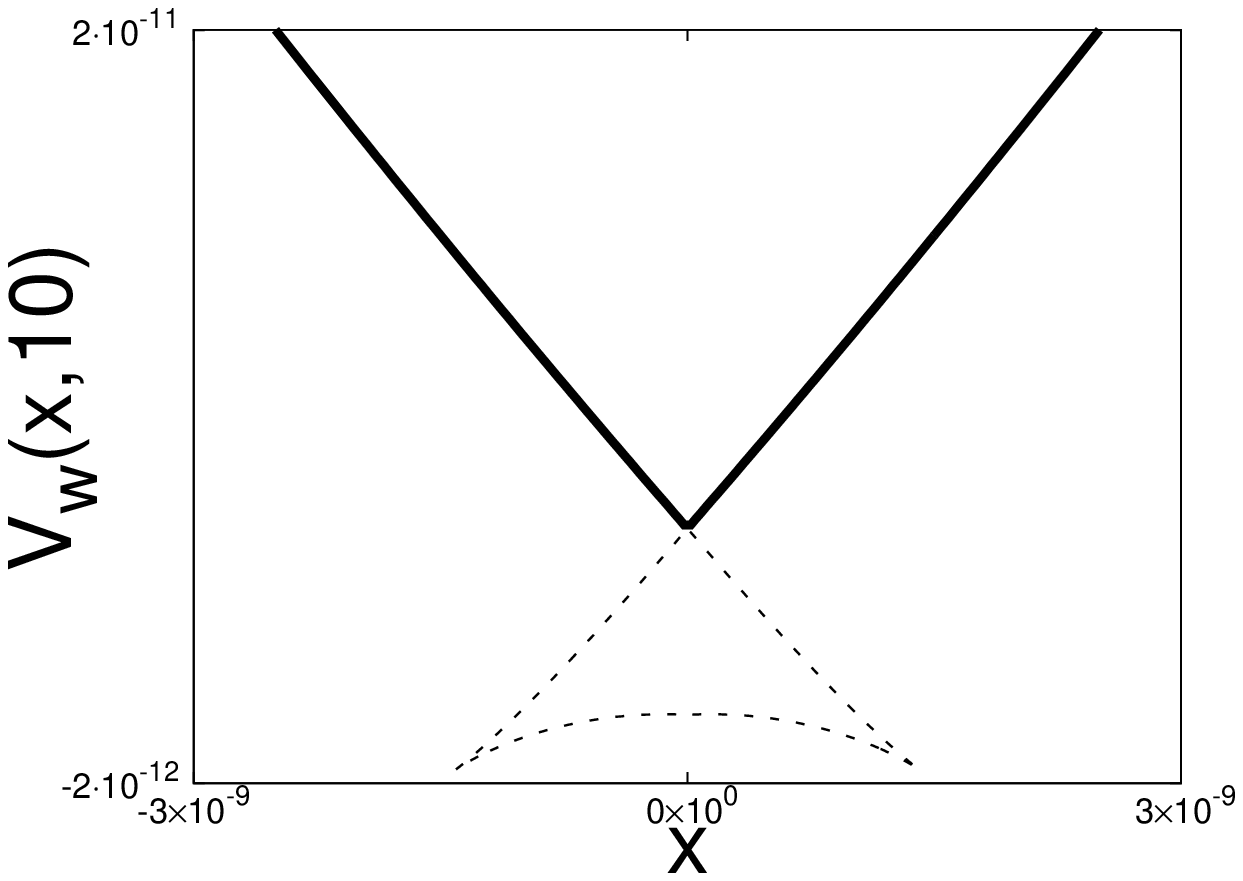}
         
         \vspace{-2mm}\hspace{5mm}(4b)\vspace{5mm}
        \end{center}
      \end{minipage}
      \hspace{2mm}
      \begin{minipage}{0.3\hsize}
        \begin{center}
          \includegraphics[width=5cm]{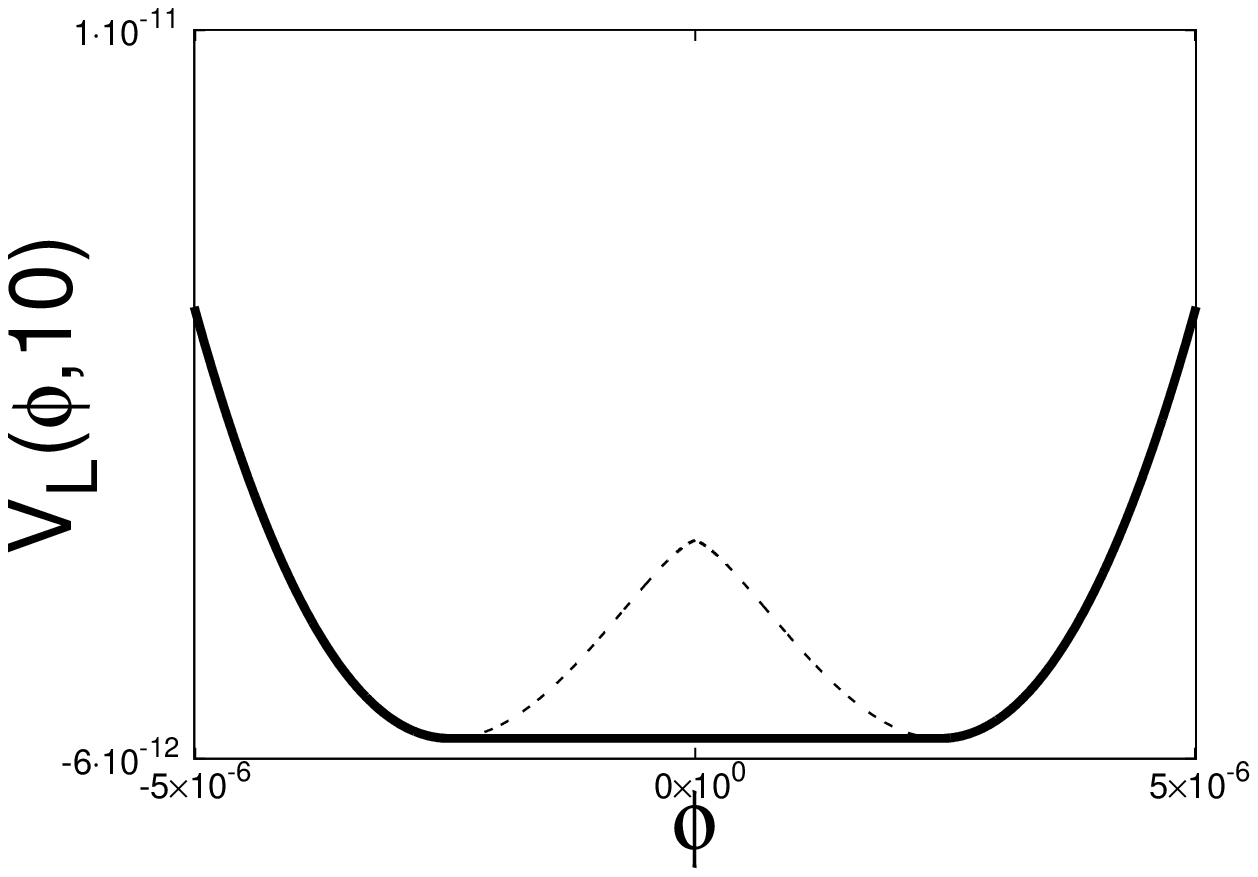}
          
          \vspace{-2mm}\hspace{5mm}(4c)\vspace{5mm}
        \end{center}
      \end{minipage}
    \end{tabular}
     \caption{Evolution of physical quantities by the weak solution 
($g=1.1 g_{\rm c}$,$~t=0.1$, $4.98$, $5.3$, $10$). (a) the mass function. (b) the Wilsonian fermion potential. (c) the Legendre effective potential.}%
  \label{fig2}
\end{center}
\end{figure}

\clearpage
\bibliographystyle{ws-procs975x65}
\bibliography{ws-pro-sample}



\end{document}